# SPIN-WAVE THEORY FOR S=1 ANTIFERROMAGNETIC ISOTROPIC CHAIN

D. V. Spirin

*V.I. Vernadskii Taurida National University, Yaltinskaya st. 4, Simferopol, 95007, Crimea, Ukraine*

*E-mail:* spirin@crimea.edu, spirin@tnu.crimea.ua

In the paper we describe the modification of spin-wave theory for one-dimensional isotropic antiferromagnet. This theory enables to obtain the energy of magnetic excitations of short wave length and correlation function in agreement with numerical studies.

The Heisenberg isotropic antiferromagnetic (AF) one-dimensional (1d) *S=1* model has been well studied. Haldane conjectured [1] that excitations in such a system possess a gap while at any temperature there is no long-range magnetic order. This statement was confirmed by numerical investigations: exact diagonalization [2], Monte-Carlo methods [3-4], real-space renormalization group [5], Lanczos technique [6]. Great progress was achieved with use of nonlinear sigma model [7].

Modified spin-wave theory (MSWT) for 1d *S=1* AF chain was formulated by Takahashi, Hirsch and Tang [8]. Later Rezende [9] extended this theory taking into account Oguchi correction. However, their results differ substantially from numerical estimations. The differences are: i) The spectrum obtained in the frameworks of MSWT has a gap $\Delta \approx 0.1J$ for $k = 0, \pi$, $k$ the wave vector, while the lower gap obtained by Nightingale and Blöte [10] is $\Delta_{k=\pi} = 0.41J$ (this value is confirmed later by numerous studies); ii) MSWT spectrum is symmetric with respect to point $k = \frac{\pi}{2}$ while numerical methods give asymmetrical spectrum with gap $\Delta_{k=0} = 2\Delta_{k=\pi}$. Providing a simple picture of a phenomena MSWT cannot predict important properties of AF *S=1* chain.

We propose modification of MSWT. We believe that in such a formulation it describes the spectrum of excitations of 1d AF isotropic chain near the point $k = \pi$.

Consider isotropic AF chain with Hamiltonian:

$$H = \sum_n \mathbf{S}_n \mathbf{S}_{n+1}, \qquad (1)$$

where summation is taken over nearest neighbors; spin of a magnetic ion is unitary: $S=1$. Let us sketch the MSWT shortly.

Using Holstein-Primakoff transformation for antiferromagnet one can obtain Hamiltonian:

$$H_{SW} = 2\sum_n \left( a_n^+ a_n - \frac{1}{2}\left( a_n^+ a_{n+1}^+ + a_n a_{n+1} \right) \right). \qquad (2)$$

We do not take into account the Oguchi correction. In the paper [9] it was used Dyson-Maleev representation, however, since Holstein–Primakoff transformation gives similar results, it does not matter which representation is applied.

To satisfy the Mermin-Wagner theorem [11], one should add such condition:

$$\frac{1}{N}\sum_n \langle S_n^z \rangle = 1 - \frac{1}{2\pi} \int_{-\pi}^{\pi} \langle a_k^+ a_k \rangle dk = 0, \qquad (3)$$

which is simply the restriction that in average the magnetic moment of one sublattice is equal to zero. Due to the condition (3) the chemical potential arises in Hamiltonian (2). It's value determines a gap in excitation energy.

One can find the spectrum of magnons:

$$\varepsilon_k = 2\sqrt{(\mu + 1 - \cos k)(\mu + 1 + \cos k)}, \qquad (4)$$

with μ the chemical potential. Eq. (4) is in agreement with that one obtained by Rezende [9] (without Oguchi correction). Solving Eq. (3) numerically we find the gap value: $\Delta_{k=0} = \Delta_{k=\pi} \approx 0.072$. The spectrum (4) is plotted at Fig. 1, *a*, (solid line); open circles correspond to results obtained in Ref. [4].

We can also find correlation functions:

$$\langle \mathbf{S}_0 \mathbf{S}_r \rangle = \begin{matrix} \frac{1}{\pi} \int_{-\pi}^{\pi} \langle a_k^+ a_{-k}^+ \rangle dk, & r \text{ odd} \\ \frac{1}{\pi} \int_{-\pi}^{\pi} \langle a_k^+ a_k \rangle dk, & r \text{ even} \end{matrix}. \qquad (5)$$

To obtain physical behavior of correlation function we take into account the four-particle term emerging from $S_0^z S_r^z$, so that

$$\langle S_0^z S_r^z \rangle \approx 1 - 2\langle a_0^+ a_0 \rangle + \langle a_0^+ a_0 \rangle\langle a_r^+ a_r \rangle \approx 0, \qquad (6)$$

due to Eq. (3). Function (5) is plotted in Fig. 1, b (solid squares). The correlations decay very slowly in comparison with the function found in numerical investigations [12]:

$$\langle \mathbf{S}_0 \mathbf{S}_r \rangle \approx const \frac{1}{\sqrt{r}} \exp\left(-\frac{r}{6.2}\right). \qquad (7)$$

Second line at Fig. 1, b (solid circles) depicts the fit of Eq.(5) by function (7).

One can conclude that MSWT underestimates the role of quantum fluctuations of the system under study. Because, in fact, MSWT consists in description of these fluctuations as waves that preserve condition (3), the slower decay of correlation function (5) with distance means that MSWT accounts only the waves of sufficiently large length.

Let us consider the spin wave of shortest wavelength $\lambda = 2$. This wave may be conceived as alternating (… 1, –1, 1 …) spin projections on the z-axis for *same* sublattice. To introduce such a wave "by hands" one may use condition:

$$\langle \mathbf{S}_r \mathbf{S}_{r+2} \rangle \approx -1, \; \mathbf{S}_r = (S_r^x, S_r^y, S_r^z), \; \mathbf{S}_{r+2} = (S_{r+2}^x, -S_{r+2}^y, -S_{r+2}^z), \qquad (8)$$

which should keep in an average. Eq. (8) means that spins of one sublattice should strongly correlate *antiferromagnetically*, if the wavelength is small. Using Holstein-Primakoff representation we have:

$$H_{SW} = 2\sum_n \left( (1+\mu)a_n^+ a_n - \frac{1}{2}(a_n^+ a_{n+1}^+ + a_n a_{n+1}) - \eta(a_n^+ a_{n+2}^+ + a_n a_{n+2}) \right), \qquad (9)$$

where we introduced second Lagrange multiplier $\eta$ to satisfy Eq. (8). The magnon energy should be found from Eq. (9) using conditions (3) and (8); Eq. (8) assumes the form:

$$\frac{1}{\pi}\int_{-\pi}^{\pi} \langle a_k^+ a_{-k}^+ \rangle dk = -1. \qquad (10)$$

The similar results may be obtained if one uses instead of restriction (8) the following:

$$\langle \mathbf{S}_r \mathbf{S}_{r+3} \rangle \approx 1, \quad \mathbf{S}_r = \left( S_r^x, S_r^y, S_r^z \right), \quad \mathbf{S}_{r+3} = \left( S_{r+3}^x, S_{r+3}^y, S_{r+3}^z \right). \qquad (11)$$

The result for excitation spectrum is presented at Fig. 2, *a*. The crosses depict a numerical result by Takahashi (see Ref. [2]), the solid line is our spectrum. For *k* lying quite near to $\pi$ the agreement is very good. We obtain a gap $\Delta_{Q=\pi} \approx 0.37 J$, and asymmetrical spectrum. Surely, we cannot describe the energy of excitations for arbitrary *k*, because the condition (10) works only near $k = \pi$.

The correlation function can be easily found. We calculate correlations from Eq.(5). For odd *r* we take the first line, where the chemical potential $\eta$ is found from condition (8), while for even *r* the second line in Eq. (5) is chosen and potential is calculated from Eq.(11). At Fig. 2, *b* this function is shown (solid triangles). Now it can be fitted by Eq. (7) very well (the fit by (7) almost coincides with obtained correlation function). For comparison we reproduce the result of MSWT from Fig. 1, *a* (solid squares).

One can conclude that without conditions like (8), (11) MSWT fails to get results in quantitative agreement with numerical ones. However, using the spin-wave "language" it is possible to describe quantitatively properties of 1d AF chain. For example, for large wavelengths one may suppose that spins of one sublattice correlate slightly *ferromagnetically* at large distance. This enables to estimate the gap value $\Delta_{k=0}$.

### ACKNOWLEGEMENTS

Author thanks Ministry of Ukraine. I also acknowledge the financial support of Ministry of Education and Science of Ukraine (grant 235/03).

Figure captions.

Fig. 1. *a* – excitation spectrum obtained with MSWT (solid line) and numerical methods (open circles; Ref. 4).

*b* – correlation function calculated with MSWT (solid circles); solid squares – exponential fit (see Eq. (7)).

Fig. 2. *a* – excitation spectrum obtained with presented version of MSWT (solid line); crosses – results of Takahashi (Ref. [2]).

*b* – correlation functions. Solid circles – the same as at Fig. 1, *b;* solid triangles – our results, which can be fit with Eq. (7) very well.

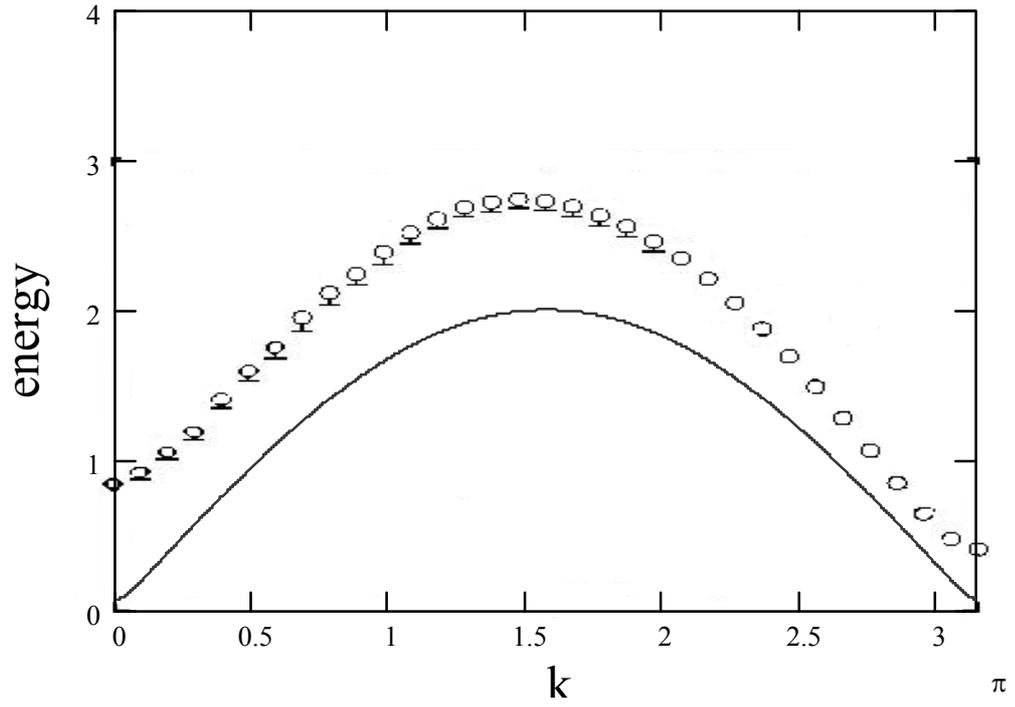

Fig. 1. *a*

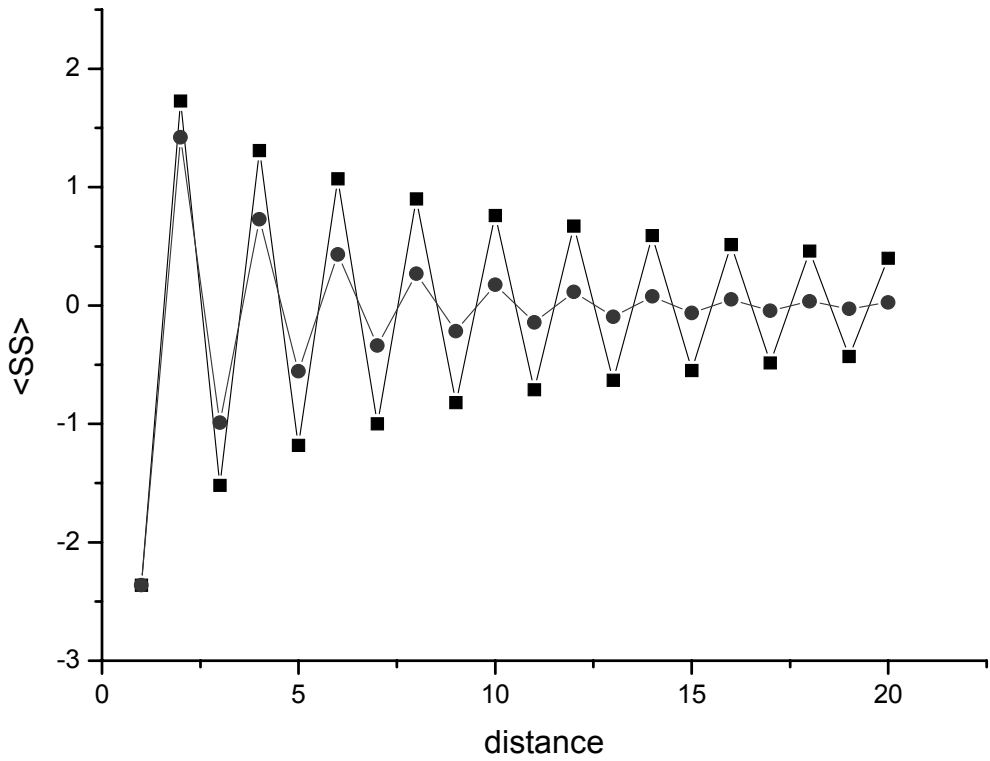

Fig. 1. *b*

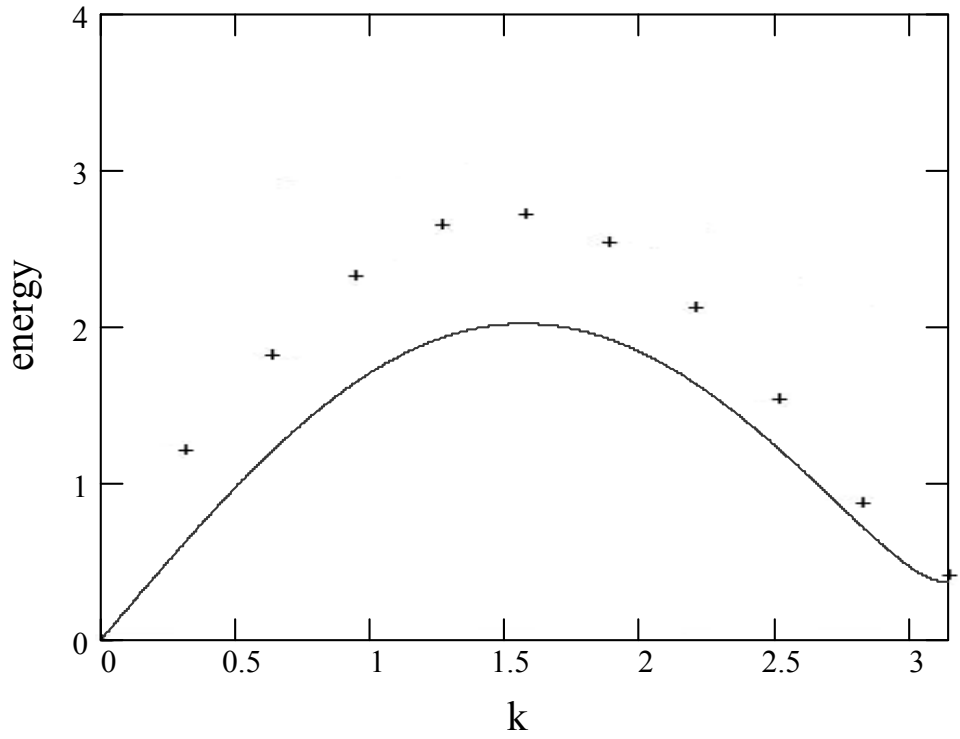

Fig. 2. *a*

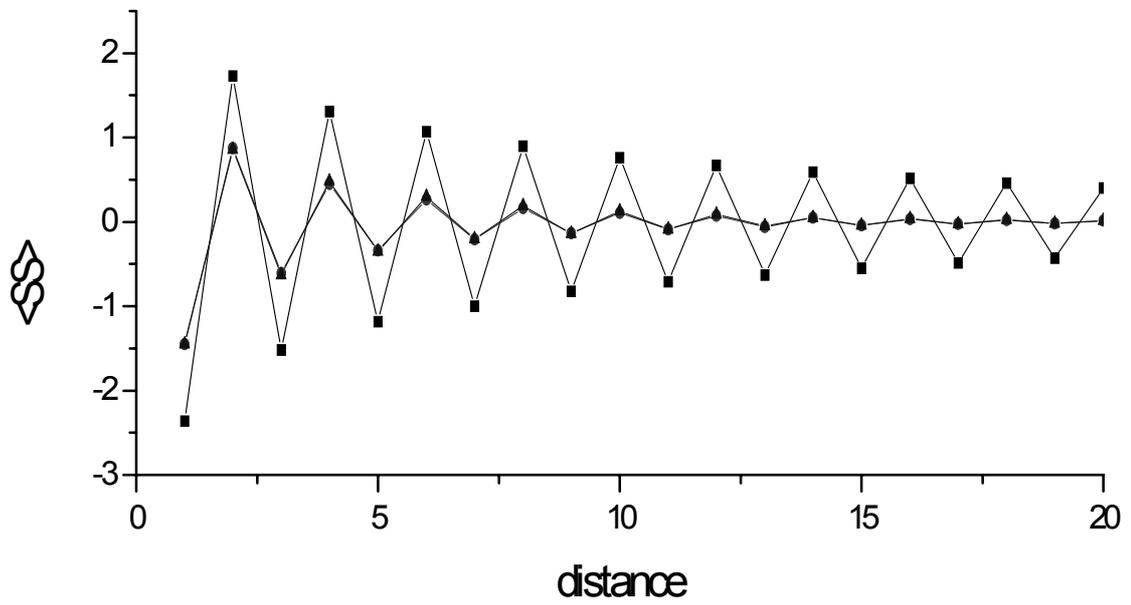

Fig. 2. *b*